\documentclass[%
rsi,
 amsmath,amssymb,
 reprint,%
]{revtex4-1}

\usepackage{graphicx}
\usepackage{placeins}
\usepackage{hyperref}
\usepackage{cleveref}
\usepackage{color,soul} 

\newcommand{\U}[2][\;]{\mathrm{#1 #2}}
\newcommand{\E}[1]{\cdot 10^{#1}}
\newcommand{\Eb}[1]{10^{#1}}

\newcommand{\dens}{m^{-3}}
\newcommand{\dg}{^{\circ}}

\newcommand{\ebd}[1]{\mathrm{10^{#1}\; m^{-3}}}
\newcommand{\ed}[1]{\mathrm{\cdot 10^{#1}\; m^{-3}}}

\newcommand{\w}{\omega}

\newcommand{\iftFac}[1]{e^{ i #1}}

\newcommand{\lb}{\left(}
\newcommand{\rb}{\right)}
\newcommand{\brk}[1]{\lb #1 \rb} 

\newcommand{\pic}[4][1.0]{
\begin{figure}[htp]
	\begin{center}
		\includegraphics[width=#1\linewidth]{#2}
		\caption{#3}
		\label{#4}
	\end{center}	
\end{figure}}

\makeatletter
\def\@email#1#2{%
 \endgroup
 \patchcmd{\titleblock@produce}
  {\frontmatter@RRAPformat}
  {\frontmatter@RRAPformat{\produce@RRAP{*#1\href{mailto:#2}{#2}}}\frontmatter@RRAPformat}
  {}{}
}%
\makeatother

\begin{document}

\title{An Innovative Heterodyne Microwave Interferometer for Plasma Density Measurements on the Madison AWAKE Prototype}

\author{Marcel Granetzny}
\email{granetzny@wisc.edu}
\author{Barret Elward}
\author{Oliver Schmitz}

\affiliation{Department of Nuclear Engineering and Engineering Physics, University of Wisconsin - Madison, Madison, Wisconsin, 53706, USA}

\date{\today}

\begin{abstract}
The Madison AWAKE Prototype (MAP) is a high-power, high-density helicon plasma experiment. The project's main goal is to develop a scalable plasma source for use in a beam-driven plasma wakefield accelerator as part of the AWAKE project. We measure the plasma density with a new heterodyne microwave interferometer that features several improvements over traditional approaches. The design uses a single microwave source combined with an upconverter to avoid frequency drift and reduce overall cost. Elliptical mirrors focus the probe beam into the plasma and guide it back to the receiver. The transmitter and receiver along with the measurement electronics are co-located in a small enclosure and are assisted by two small mirrors on the opposite side of MAP. Both halves of the system move independently on computer-controlled motion platforms. This setup enables fast repositioning of the interferometer to measure at any axial location despite the magnets, wiring and structural supports that would block movement of a waveguide-based system. A high-speed, high-precision mixed signal circuit and FPGA analyze the probe signal directly in the enclosure which obviates the need for a digitizer or oscilloscope. The interferometer resolves phase shifts down to one hundredth of a fringe, resulting in a line-averaged resolution of $1.5\ed{17}$. The system provides a real-time measurement every $5\U{\mu s}$ up into the mid $\ebd{19}$ density range with a noise level of $1.0\ed{17}$. 
\end{abstract}

\keywords{keywords here}

\maketitle

\section{Introduction}


 The Madison AWAKE Prototype (MAP) is a plasma development platform that has been built as part of CERN's beam-driven plasma wakefield accelerator project AWAKE\cite{Caldwell2016,gschwendtner2016awake}. MAP uses a dual helicon\cite{Boswell1984,Chen2015} antenna setup with up to 20 kW of RF power to create plasmas with densities up into the $\ebd{20}$ range in a highly uniform magnetic field. A detailed description of the MAP project and its main results is given in \cite{granetzny2025MAP}.\\

\begin{figure}[htp]
	\begin{center}
		\includegraphics[width=1.0\linewidth]{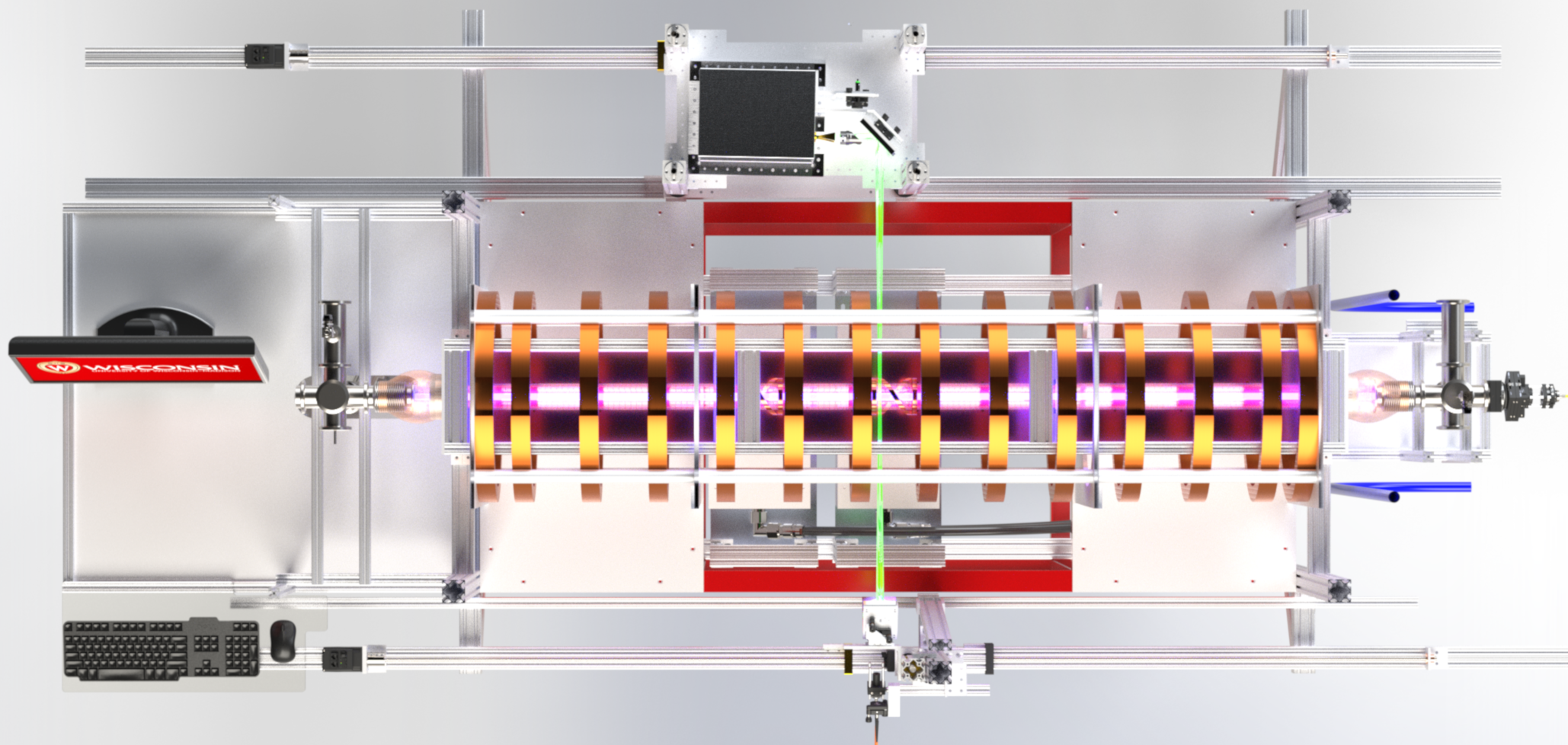}
		\caption{Simplified CAD render showing a top view of MAP with the 2.6 m long vacuum vessel and the microwave interferometer. The probe beam is sent out from the enclosure on the top and passes through the plasma as indicated by the green beam. The beam is sent back to the enclosure by the mirror system on the bottom. Both parts of the interferometer can be moved horizontally and vertically using independent computer-controlled motion platforms.}
		\label{fig:MAPTop}
	\end{center}	
\end{figure}

MAP's success depends on fielding a reliable and precise density diagnostic to measure and optimize the plasma density profiles for practical use in a wakefield accelerator. To this end, we have built a 105 GHz heterodyne microwave interferometer. \Cref{fig:MAPTop} shows a top view of MAP and the interferometer setup. The interferometer sends out a microwave beam, which passes through the plasma and is guided back into the receiver. This probe beam experiences a phase shift that depends on the plasma density along the beam path. This phase delay can be analyzed to provide line-integrated density measurements. Computer-controlled motion platforms allow the entire system to move vertically and horizontally to enable density measurements anywhere in the MAP plasma.\\

\label{sec:hetInfIntro}
Heterodyne interferometry is a common diagnostic used to measure electron densities in plasmas. An extensive overview of the subject in the greater context of millimeter wave based plasma measurements is provided in \cite{hartfuss1997heterodyne}. Our discussion here will focus on the core physical principle and its consequences for our diagnostic design.\\

In an interferometric plasma density measurement, an O-mode wave, with angular frequency $\w$, is sent into the plasma. This probe beam will undergo a phase shift $\Delta \phi$ depending on the plasma density $n_e$ along the beam path as \cite{Hutchinson2002}[pp. 112-116]

\begin{align}
\label{eq:infBasic}
\Delta \phi &= - \frac{\w}{2 c n_c}\int n_e \,dl \quad\text{with}\quad n_c = \frac{\w^2 m_e \epsilon_0}{e^2},
\end{align}

where $n_c$ is the cutoff density and we have assumed $n_e \ll n_c$. Under this condition the plasma's index of refraction ($\sqrt{1-n_e/n_c}$) is close to 1, and the probe beam does not undergo significant refraction while passing through the plasma. The phase shift is measured by interfering the probe beam with a reference beam that travels a fixed path outside the plasma.\\

In a heterodyne setup, the plasma and reference beams operate at slightly different frequencies with electric field strengths

\begin{align}
E_p &= |E_p(t)|\iftFac{\brk{\w_p t + \phi(t)}}\quad\text{and}\quad
E_r = |E_r|\iftFac{\w_r t},
\end{align}

where $\w_p$ and $\w_r$ are the probe and reference beam frequencies, respectively. The plasma affects the probe beam and introduces a time-dependent amplitude and phase. The signal strength at the detector $S = \left| E_p + E_r \right|^2$ is 

\begin{align}
\label{eq:sig}
S(t) &= |E_p(t)|^2+|E_r|^2 + 2|E_p(t)||E_r| \cos \brk{2\pi f t + \phi(t)},
\end{align}

where $f= (\w_p - \w_r)/2\pi$ is commonly called the modulation frequency. In a practical setup, we have $\w_p > \w_r$ and $2\pi f \ll \w_r$, such that the modulation frequency raises the probe beam frequency only slightly above the reference beam frequency. The latter is often referred to as the carrier frequency.\\

If the signal amplitude rate of change is slow compared to the modulation frequency, we can use an analog or digital lowpass filter to extract only the last term in \cref{eq:sig}. An example of such a signal with time-varying phase shift is shown in \cref{fig:hetInfFastShift}. The top panel shows the AC part of the signal in blue, with a frequency that is modulated by a time-dependent phase shift (green). This phase shift either delays or advances the zero crossings of the measurement signal relative to an unshifted signal at the modulation frequency (orange). The bottom panel shows a logic conversion of the signal, indicating whether the signal is above or below zero Volts. We can then measure the time delay $\Delta t$ between the measurement signal's zero crossings and the expected zero crossing for a signal at the modulation frequency. At any moment the phase is proportional to the time delay $\Delta T$ as $\phi(t) = 2\pi f \Delta T(t)$.\\

\pic[1.0]{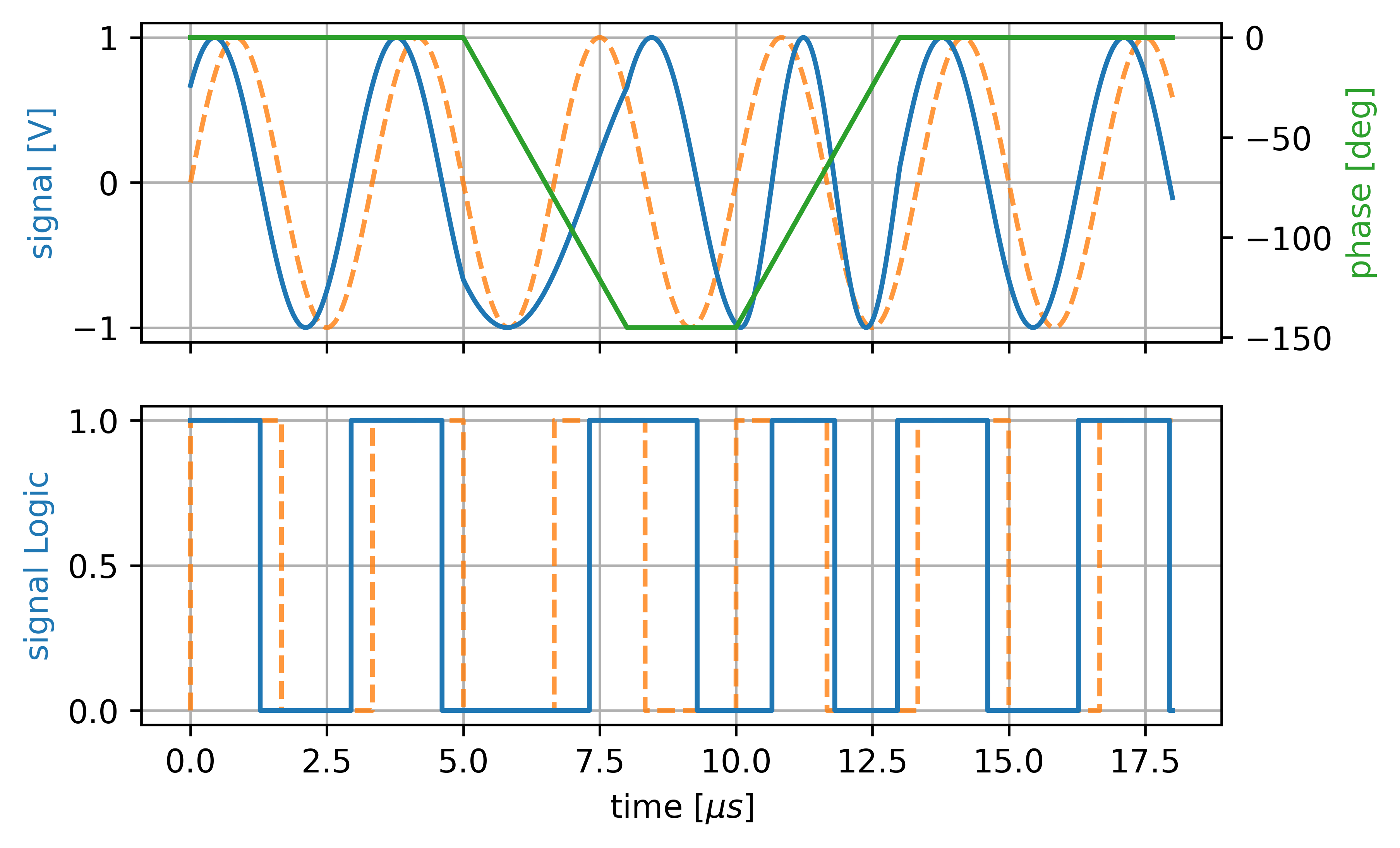}{Example of heterodyne interferometer AC signal with $f = 1\U{MHz}$ and a rapid phase shift. Top: The measurement signal with phase shift (blue), the expected signal without phase shift (orange dotted), and the phase shift itself (green). Bottom: Conversion of the measurement signal into logic (blue), indicating whether the signal is above or below 0 V, along with the logic reference (orange).}{fig:hetInfFastShift}

Unlike a homodyne system, which uses $\w_p = \w_r$, a heterodyne system decouples the phase measurement from an amplitude measurement and can therefore measure density even in cases of strong beam attenuation. It further resolves any ambiguity in the direction of the phase shift since an increase in phase leads to a decrease in the time delay between zero crossings and vice versa. A full period ($1/f$) delay equals a $2\pi$ phase shift and is commonly referred to as a fringe with anything smaller being referred to as a fractional fringe.\\

\section{The MAP Interferometer System}
\subsection{Design Considerations and Traditional Solutions}
\label{ssec:designConsiderations}

\paragraph{Choice of Frequency}

Typical sources for plasma interferometry operate at far-infrared (FIR) or microwave frequencies. The most common FIR source is a  CO\textsubscript{2} laser at $10.6 \U{\mu m}$ or 28.3 Thz. The advantages of an FIR system are minimal refraction in the plasma and very high cut-off density $n_c$. According to \cref{eq:infBasic}, a CO\textsubscript{2} laser experiences cutoff at $9.9\E{24}\U{\dens}$, four orders of magnitude above the AWAKE target density. However, the resulting phase shifts are very small. For example, assuming a line-averaged density of $\Eb{19}\U{\dens}$ across the 52 mm wide MAP vacuum chamber leads, according to \cref{eq:infBasic}, to a phase shift of just $0.9\dg$. An FIR system therefore requires very fast data acquisition or suffers from low-density resolution. In addition, an FIR system is extremely sensitive to vibrations. For example, a change in the beam path length of just $27 \U{nm}$ leads to a phase shift of $0.9\dg$ and is therefore indistinguishable from a line-averaged plasma density of $\Eb{19}\U{\dens}$. The other common frequency choice is the microwave band with wavelength in the millimeter range and frequencies of a few hundred GHz, which yield significantly higher density resolution and vibration resistance. Using a 2.9 mm source at 105 GHz, a $\Eb{19}\U{\dens}$ plasma in MAP would cause a phase shift of $240\dg$. Even strong vibrations which might change the beam length by $100 \U{\mu m}$ would only result in a $12\dg$ phase shift, equivalent to a line averaged density of $5\E{17}\U{\dens}$. The main disadvantage of microwave systems is their relatively low cut-off frequency. At 105 GHz the cutoff density is $1.4\E{20}\U{\dens}$.\\

In between the microwave and FIR bands lies what is commonly called the Terahertz regime, which combines the benefits of both. For example, a 700 GHz system has a cutoff density of $6\E{21}\U{\dens}$, which is an order of magnitude above the AWAKE target density. At the same time, such a system offers good density resolution and is not significantly affected by vibrations. The main disadvantage is the cost of contemporary Terahertz sources, which are still in the early phases of commercialization. The other significant difference between FIR on the one hand and microwave or Terahertz systems on the other is achievable spatial resolution. Just like optical sources, FIR lasers have low beam divergence and their output can be well collimated down to spot sizes under one millimeter. In contrast, microwave and Terahertz sources have large beam divergences on the order of ten degrees.\\

\paragraph{Frequency Stability}

The next challenge is to create a constant frequency difference between the probe and reference beam to avoid what is known as frequency drift. Frequency drift results in any measurements being inaccurate as the modulation frequency in \cref{eq:sig} becomes time-dependent. One attempt at a solution uses two highly stable sources, one for the probe and one for the reference beam. However, the needed accuracy would need to be phenomenal. For example, for a 100 GHz carrier frequency with a 1 MHz modulation and an acceptable frequency drift of 1\%, we would need a frequency stability of 50 ppb in each source. Such highly accurate sources are not available for practical purposes. Frequency drift is therefore usually solved by employing two sources with variable frequency in combination with a highly accurate master oscillator. The latter outputs a signal at the desired modulation frequency and drives a phase-locked loop (PLL) which in turn adjusts the frequency of one source\cite{Forest1990,Smith2018} to stabilize the modulation frequency. This setup ensures that the modulation frequency is as stable as the master oscillator. The disadvantage is the relatively high cost associated with using one or two variable frequency sources, a master oscillator and a PLL.\\

\paragraph{Fractional Fringe Resolution}
Our final consideration involves speed and the needed resolution of the data acquisition system. A typical interferometer signal in the setup presented later might have a signal amplitude of a few hundred $\U{\mu V}$. As explained in \cref{sec:hetInfIntro}, the achievable phase resolution depends on the available time resolution. At a 1 MHz modulation frequency, resolving a hundredths of a fringe requires digitizing the signal at 100 MHz, or 100 Mega samples per second (MSPS). A high-end DAQ system, such as the D-tAcq ACQ480FMC\cite{dtacq}, can achieve 80 MSPS, with a $300 \U{\mu V}$ resolution, albeit with a $\pm 5 \U{mV}$ measurement error. Such a system is therefore not quite accurate enough to capture the small signals involved and would need to be augmented by a low-noise amplifier.\\

\subsection{Overview of the MAP Interferometer System}

\begin{figure*}[htp]
	\begin{center}
		\includegraphics[width=0.8\linewidth]{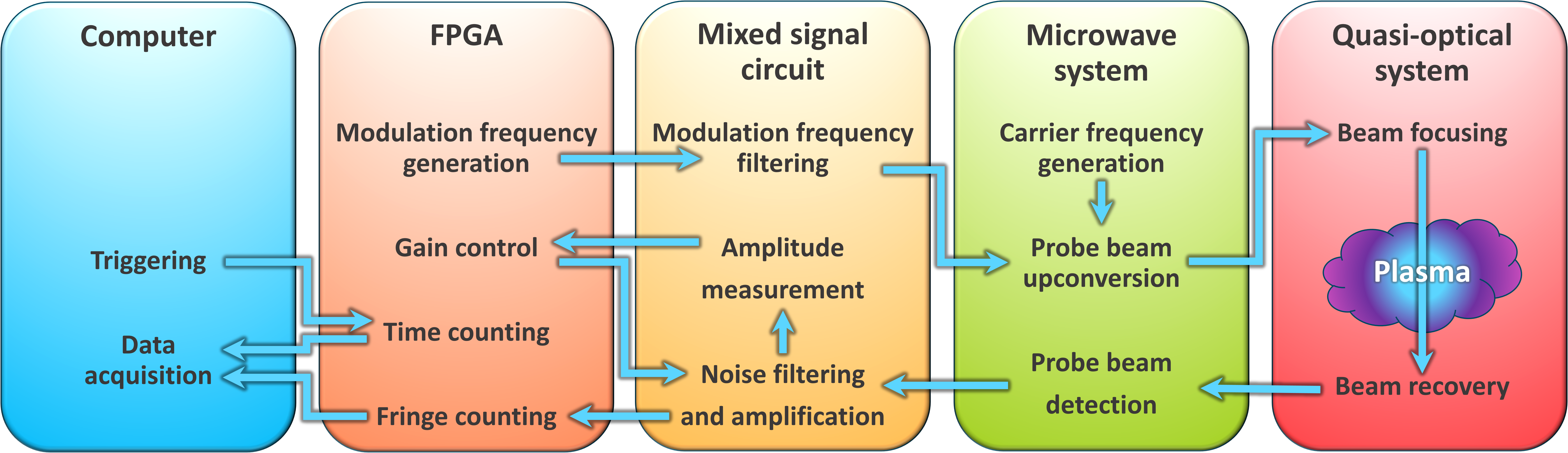}
		\caption{Flow chart of the density measurement process.}
		\label{fig:hetInfOverview}
	\end{center}	
\end{figure*}

In light of the discussion in \cref{ssec:designConsiderations}, the MAP interferometer uses a 105 GHz microwave source to provide high density resolution and reduce the impact of vibrations enough to make the whole setup portable. This allows us to position the interferometer at different axial and height positions to measure 2D density profiles. Our design uses several innovative techniques to deal with the aforementioned issues of frequency drift, data acquisition requirements, and beam divergence. An overview of the system is given in \cref{fig:hetInfOverview} in form of a flow chart.

\begin{figure*}[htp]
	\begin{center}
		\includegraphics[width=1.0\linewidth]{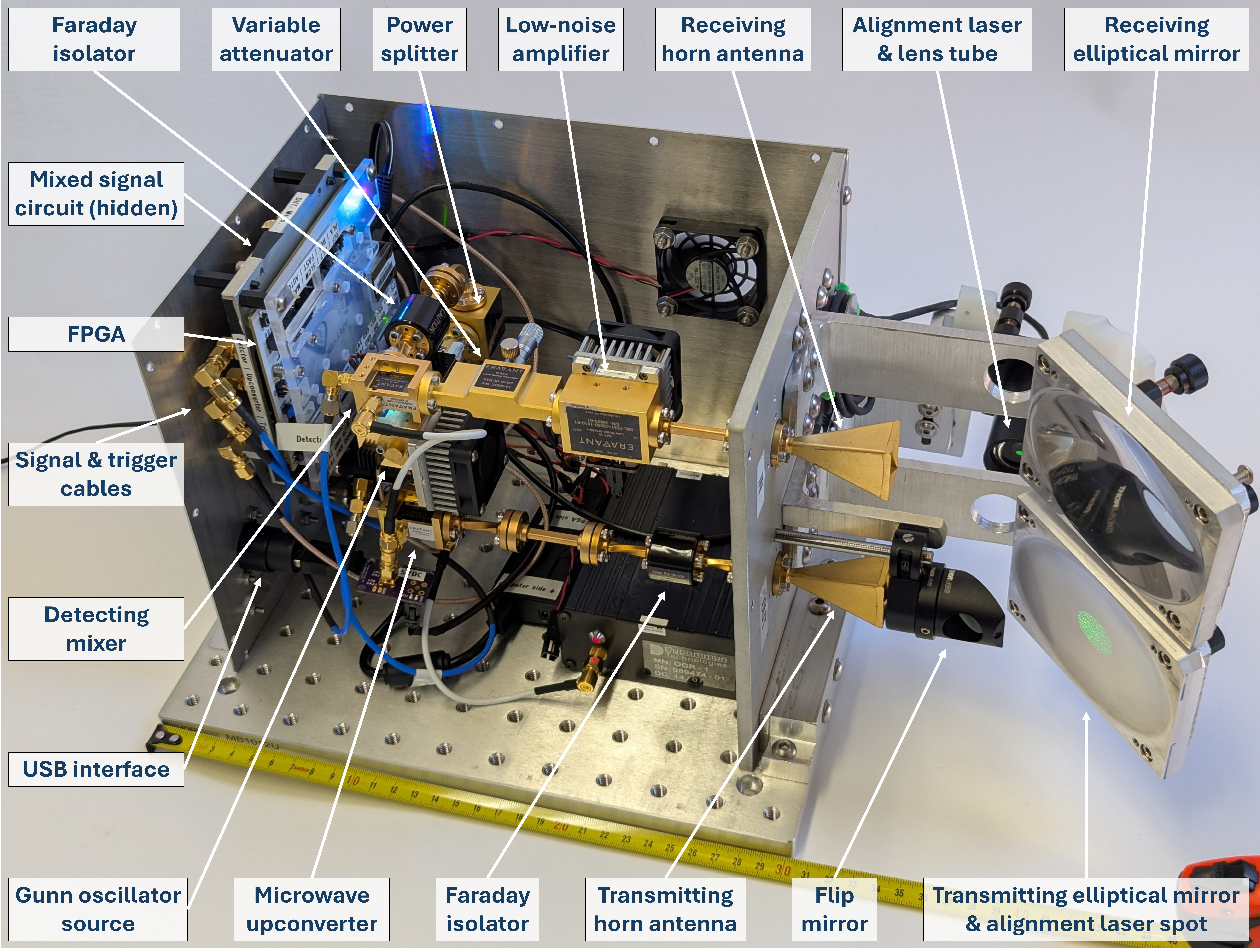}
		\caption{Photograph of the entire interferometer system, except for two of the mirrors, with tape measure for scale. The enclosure's top and front side panels have been removed to show the FPGA, microwave components, connecting cables, power supplies, alignment laser, and two elliptical mirrors. The configuration shown is the one for alignment in which the green laser matches the microwave beam's divergence and focal point distance to the first mirror. During measurement, the small flip mirror is rotated 90 degrees up to allow the microwave beam to pass into the focusing elliptical mirror unobstructed.}
		\label{fig:hetInfOverviewReal}
	\end{center}	
\end{figure*}

Starting with the orange panel in \cref{fig:hetInfOverview}, an FPGA generates a digital modulation signal at 1 MHz. The mixed signal circuit converts this signal into a pure 1 MHz sine wave and sends it to the microwave system. The microwave system employs a single 105 GHz fixed frequency source whose output is split it into a reference and a probe beam. The probe beam and the 1 MHz modulation signal are combined in an upconverter that shifts the probe beam to 105.001 GHz. This beam is sent through a horn antenna and into a quasi-optical system of flat and elliptical mirrors that guides the beam through the plasma and back into a receiving antenna while keeping it focused throughout. The return probe beam is combined in a microwave mixer whose output is the modulation signal at 1 MHz, including plasma-induced phase shifts. This signal is amplified and noise filtered on the mixed signal circuit before a high-speed comparator detects zero crossings and converts the measurement signal into a train of logic pulses as exemplified in the bottom panel of \cref{fig:hetInfFastShift}. The logic signal is analyzed by an FPGA every 10 ns, which enables a resolution of one-hundredth of a fringe, or $3.6\dg$, equivalent to a line-averaged density resolution of $1.5\E{17}\U{\dens}$ across MAP's $52\U{mm}$ diameter plasma. The FPGA keeps track of the total fractional fringe count along with an internal time stamp. Both are sent to a computer through a USB interface which enables real-time density measurements every $5\U{\mu s}$. This system obviates the need for a high-speed digitizer, adjustable frequency source, master oscillator, and phase-locked loop found in traditional heterodyne interferometer systems. At the same time, the electronics and optics are designed to be frequency agnostic. The existing 105 GHz system can be swapped out for a Terahertz or FIR system without modification of any other part of the interferometer setup. \Cref{fig:hetInfOverviewReal} shows a photograph of the source side of the fully assembled system. We will describe the details of the individual system components in the following sections.

\subsection{FPGA Logic Circuit and Computer Interface}
\label{sec:fpga}

Our FPGA design uses the Digilent Arty A7-35T board and has been programmed in Verilog to execute the following workflow. First, the 100 MHz internal clock is down-sampled to a 1 MHz digital reference signal and sent to the mixed signal circuit described later in \cref{sec:circuit}. At the same time, the FPGA receives a digital modulation signal, including any plasma-induced phase shifts from the mixed signal circuit. We will call this signal the plasma signal. The algorithm then measures the timing difference between positive logic edge transitions of the reference and plasma signals by comparing the signals' logic states every clock cycle at 100 MHz. The timing difference, or fractional fringe count is stored in an internal register. The algorithm keeps track of corresponding edges in the reference and plasma signals and can therefore track them across any number of full fringes. In addition, every time the reference signal undergoes a positive edge transition an internal time counter is incremented. Fringe count and time stamp data along with cyclic redundancy check bytes\cite{houghton1997cyclic} are sent to the control computer at 12 Megabits per second through a USB-UART bridge. This system provides a full density measurement every $5\U{\mu s}$ in steady-state streaming operation. The time counter can be reset by a command from the control computer or analog trigger signal, to synchronize the internal time stamp with the global shot time in pulsed MAP discharges. Lastly, the FPGA receives two signals from a window comparator setup on the mixed signal circuit that evaluate the analog plasma signal's amplitude. If the amplitude is out of bounds, the FPGA sends control pulses to the mixed signal circuit to change the amplifier gains as explained later in \cref{sec:circuit}.


\subsection{Mixed Signal Circuitry}
\label{sec:circuit}

The interferometer uses a high-speed, high-precision, mixed signal circuit to interface between the FPGA and the microwave system described later in \cref{sec:microwave}. This circuit generates the sinusoidal modulation signal, filters and amplifies the plasma signal, converts the plasma signal into logic pulses, and sets the signal amplitude.\\

\begin{figure*}[htp]
	\begin{center}
		\includegraphics[width=1.0\linewidth]{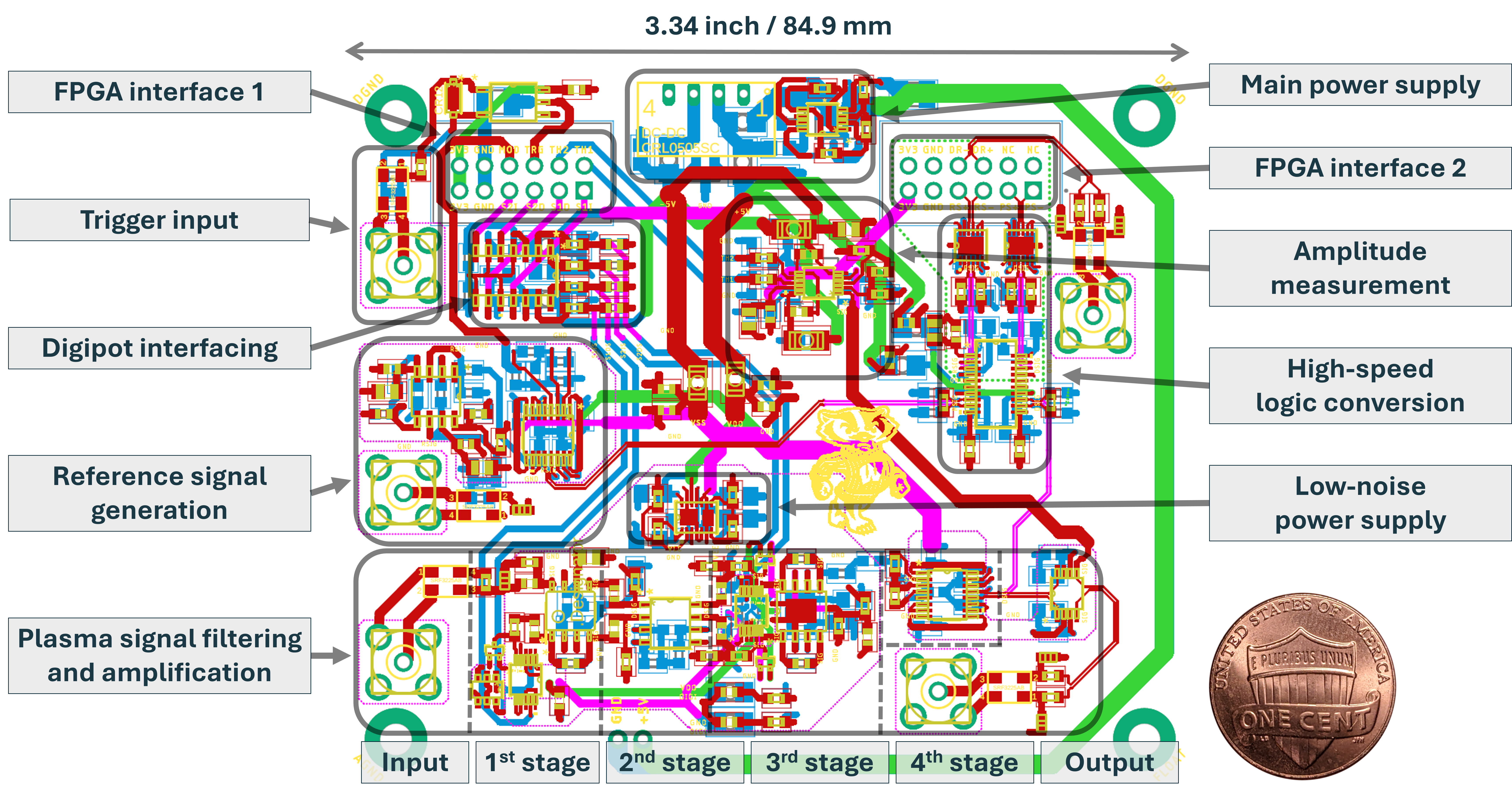}
		\caption{View of the signal and power carrying layers of the mixed signal circuit board with a penny for scale. To improve signal integrity, different kinds of signals are separated into different layers and separated by two continuous ground planes. Analog signals are on the top layer (red) and rarely the first inner layer (purple). Digital signals are on the bottom layers (blue) and rarely the second inner layer (green). Power rails, supplying seven different voltage levels, are shown in purple and green.}
		\label{fig:PCBEagle}
	\end{center}	
\end{figure*}

\paragraph{PCB Layout}
We implemented this circuit on a six-layer printed circuit board (PCB) to optimize performance and reduce noise. An annotated PCB layout, omitting ground planes, is shown in \cref{fig:PCBEagle}. The top layer (red) is used to carry and process analog signals. Digital control signals are confined to the bottom layer (blue). The different power rails are primarily confined to the inner two layers (pink and green). Layers two and five are used as continuous ground planes. This six-layer design allows us to separate digital from analog and power signals, all separated from each other by two dedicated ground planes. All measurement signals traveling farther than a few millimeters are sent down differential pairs acting as transmission lines with 100 $\Omega$ impedance. All active components use multiple decoupling capacitors and the probe signal processing elements run on their own low-noise power supply. The ground planes are stitched together using a few hundred vias.\\

\paragraph{Modulation frequency generation}

As described in \cref{sec:fpga}, the FPGA provides a 1 MHz reference frequency. However, this signal is digital, whereas the microwave components need a pure 1 MHz sinusoidal signal. We create this signal with a  4th-order bandpass filter that rejects the DC and all but the 1 MHz components. The microwave circuit upconverter is driven by a voltage-clamping operational amplifier with sufficient current to allow for a 50 $\Omega$ termination of the outgoing coaxial cable.\\

\paragraph{Plasma signal processing}

Due to attenuation in the microwave components and plasma, the measurement signal has a variable amplitude of just a few hundred microvolts. In addition, the interferometer is necessarily located close to the dual helicon antenna setup, which uses up to 20 kW of RF power at 13.56 MHz. Even though MAP has a highly effective Faraday screen that reduces measured free space RF emission amplitudes by up to 49 dB\cite{granetzny2025MAP}, this presents a potentially enormous noise signal at a frequency only one decade removed from the modulation frequency. Further, the microwave detector has a high output impedance and low current sourcing capability. We use a carefully designed high-precision analog signal processing chain that addresses all of those issues and recovers a clean measurement signal.\\ 

The microwave mixer's high output impedance precludes AC coupling and the use of a noise-reducing 50 $\Omega$ coaxial termination where the plasma signal enters the PCB. Our first stage is therefore a high input impedance operational amplifier (op-amp) in a modified non-inverting amplifier configuration. The standard non-inverting scheme is modified with capacitors in the feedback path that turn the op-amp circuit into a follower at high and low frequencies. The result is a band-amplifier centered around 1 MHz. The design uses a digital potentiometer in the feedback loop to allow for a variable gain. To improve noise immunity, this stage also converts the signal from single-ended to differential.\\

The second stage consists of a 7th-order lowpass filter with a 2.3 MHz corner frequency. This component eliminates practically all remaining RF noise by attenuating frequencies at 13.56 MHz by 70 dB relative to 1 MHz.\\

A third stage is used to bring the signal into the $1-2 \U{V_p}$ range by combining a fully differential, voltage-limiting amplifier with double digital potentiometer in a dual tracking configuration. This allows us to keep the signal fully differential, while voltage clamping ensures that the output signal amplitude does not exceed the maximum input voltage level of the downstream comparator.\\

A final, fourth stage consists of a wide passband filter centered around 1 MHz that eliminates any remaining RF noise, DC offsets and lower frequency power supply noise. Due to the digital potentiometers, this four-stage signal processing chain has a variable gain ranging from approximately 2,000 to 150,000 at 1 MHz.\\

The filtered and amplified differential plasma signal is sent to a line driver and into a high-speed comparator. The line driver provides a signal for readout by an oscilloscope, which is useful when aligning the interferometer's quasi-optical system described later in \cref{sec:quasioptics}. This component also sends the signal into a window comparator which we will describe shortly. Meanwhile, the high-speed comparator detects the signal's zero crossings and creates a logic signal for the FPGA, which requires fast and bounce-free signal transitions for fringe counting. Our setup therefore uses a comparator with a rise time under 1 ns with switching hysteresis. The latter ensures bounce-free transitions while the former allows the system to operate with FPGA clock frequencies up to 1 GHz in case finer density resolution is needed in the future.\\

\begin{figure}[htp]
	\begin{center}
		\includegraphics[width=1.0\linewidth]{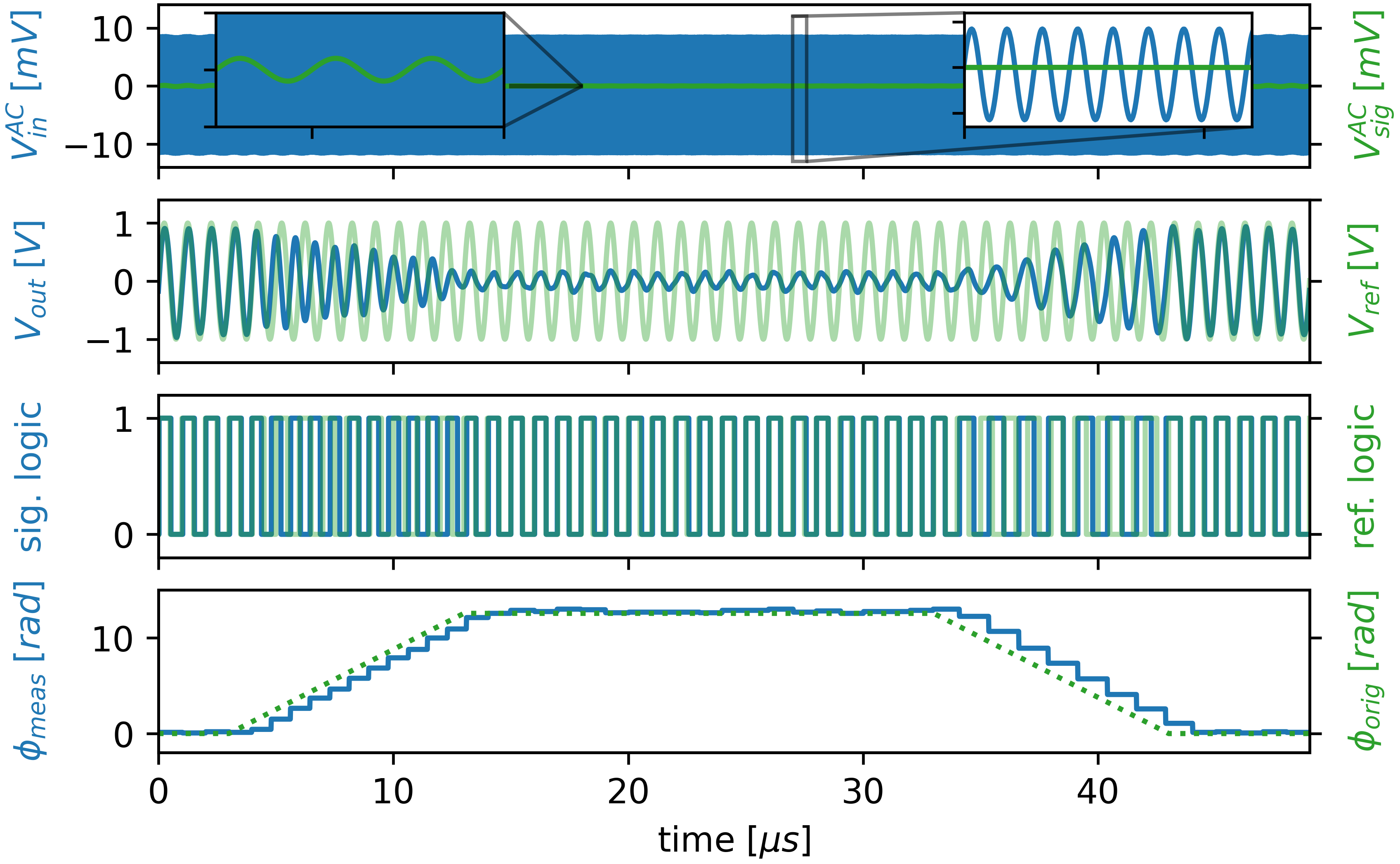}
		\caption{Results of an LTSpice simulation for amplifier and filter chain on the mixed signal circuit. Top panel: Full incoming signal, including RF noise (blue) and the part of the signal that contains density information (green). Second panel: Signal after noise filtering and amplification (blue) along with the reference signal (green). Third panel: Logic conversion of measurement and logic signals. Bottom panel: Phase shift calculation from measurement signal and comparison to applied phase shift.}
		\label{fig:circuitInputSim}
	\end{center}	
\end{figure}

An LTSpice\cite{LTSpice} simulation of the signal processing circuit described above is shown in \cref{fig:circuitInputSim}. The top panel shows the AC part of the incoming plasma signal (blue). In this simulation, the useful signal (green) has an amplitude of $100 \U{\mu V p}$ and experiences a fast phase shift of $4\pi \; (720\dg)$ in  $10 \U{\mu s}$. The equivalent frequency differential for such a rapid phase shift is 200 kHz, such that our measurement signal would fall into the 0.8 to 1.2 MHz window during this discharge. In addition to this plasma signal, we impose a hundred times stronger RF noise signal with an amplitude of $10 \U{m V_p}$ as a worst-case scenario. In addition, we assume a white noise signal with an amplitude of $20 \U{\mu V_p}$ and that the signal amplitude is reduced by a factor of five at maximum phase shift due to attenuation in the plasma. The second panel shows the signal at the end of the filter and amplifier chain in blue compared to a 1 MHz steady reference signal in green. Note that the scale has changed from mV to Volts and that all RF noise has been filtered out. The third panel shows the plasma and reference signals after logic conversion. The final panel shows the reconstructed phase shift (blue) based on fringe counting using the FPGA algorithm described in \cref{sec:fpga}. The reconstructed phase shift closely tracks the imposed phase shift with a time delay of $1 \U{\mu s}$.\\ 

\paragraph{Amplifier feedback control}
The plasma signal amplitude has to stay within the voltage limits and gain-bandwidth products of the different amplifier chain components. At the same time, a high amplitude results in fast zero crossings which are beneficial to logic conversion by the high-speed comparator. We achieve both by sending the final plasma signal into a window comparator that allows us to stabilize the amplitude between a lower and upper voltage threshold. The window comparator's output is analyzed by the FPGA described in \cref{sec:fpga}, which in turn sends control pulses to the digital potentiometers to adjust the amplifier gains of the first and third stages automatically and continuously. This allows us to vary the overall amplifier gain from approximately 2,000 to 150,000. Many digital potentiometers are `glitching' when changing between tab positions, meaning they appear temporarily open. Our setup requires continuous signal processing, which is why it uses a glitch-free component. However, different settings in any potentiometer result in small changes of the component's parasitic capacitance which in turn changes the impedance in the op-amp feedback loop. The result is a position-dependent phase shift of the signal that would be indistinguishable from a change in plasma density. To avoid this problem, we programmed the FPGA to stop auto-adjusting the amplifier gains for the remainder of each shot after receiving a trigger pulse from the control computer. Automated gain adjustment is therefore only used before each plasma pulse to compensate for location-dependent attenuation factors such as variations in the vacuum vessel, Faraday screen, and possible beam path misalignments which can change when the interferometer is moved to a new position as described later in \cref{sec:motionSystem}. This procedure ensures the amplifier chain gain is set to its optimal value before each shot. 

\subsection{Microwave System}
\label{sec:microwave}

\Cref{fig:hetInfOverviewReal} provides an overview of the microwave system and it's connections to the FPGA and quasi-optics. A fixed-frequency Gunn oscillator is used to generate a 13.7 dBm signal at 105 GHz. This signal is sent through a Faraday isolator and a 6 dB attenuator. The latter is necessary to stay below the damage limits of the following components. A 50:50 power divider splits the microwave signal into a reference beam and a probe beam, both of which pass through Faraday isolators to prevent leakage from the reference into the probe beam. The reference beam is sent to the detecting mixer. Meanwhile, the probe beam enters an upconverter that shifts it to 105.001 GHz using the modulation signal from the mixed circuit board described in \cref{sec:circuit}. This upshifted probe beam passes through a final Faraday isolator and exits through the lower rectangular horn antenna into the mirror system described later in \cref{sec:quasioptics}. This system guides the beam through the plasma and into the upper horn antenna. After passing through a 35 dB low noise amplifier, the probe beam is sent into an adjustable attenuator before combining with the reference beam in the microwave mixer. This amplifier-attunator combination allows us to set the signal amplitude within the range allowed by the mixer, without prior knowledge of the exact beam attenuation factor. The mixer multiplies the reference beam with the probe beam as described mathematically by \cref{eq:sig}. The resulting signal is centered at the modulation frequency, carries the plasma-induced phase shifts and is sent to the mixed signal circuit for processing as described in \cref{sec:circuit}.

\subsection{Quasioptical System and Alignment}
\label{sec:quasioptics}

A significant challenge when using a microwave setup is the beam's high divergence. The horn antenna shown in \cref{fig:hetInfOverviewReal} has a beam width of $7.7\dg$ at 105 GHz. The distance from the interferometer to the plasma is about 54 cm and would result in a 76 mm beam diameter at the plasma, therefore exceeding the vacuum vessel diameter. Achieving a meaningful spatial resolution therefore requires focusing the beam. Unfortunately, microwaves cannot be focused with glasses used for optical frequencies. On the other hand, certain plastics can be used to make microwave lenses but are useless at optical frequencies. The latter is important if we send an alignment laser down the same path as the microwave beam. An ideal optical system for our purposes needs to work at optical and microwave frequencies. This is possible by constructing the entire system out of mirrors, whose optical properties are frequency-independent up into the ultra-violet regime.\\

When designing laser-based optical systems, the beam divergence is usually small enough to assume parallel rays and work in the paraxial approximation. Any beam transfer path can then be constructed with planar mirrors and focusing can be achieved by a single parabolic mirror at the end. In contrast, our microwave source requires elliptical mirrors that can focus the highly divergent beam from one focal point to another, for example from the microwave source into the plasma core. We show in \cref{app:MirrorDesign} how to design an elliptical mirror given source and target focal lengths and an arbitrary reflection angle.\\

\begin{figure*}[htp]
	\begin{center}
		\includegraphics[width=1.0\linewidth]{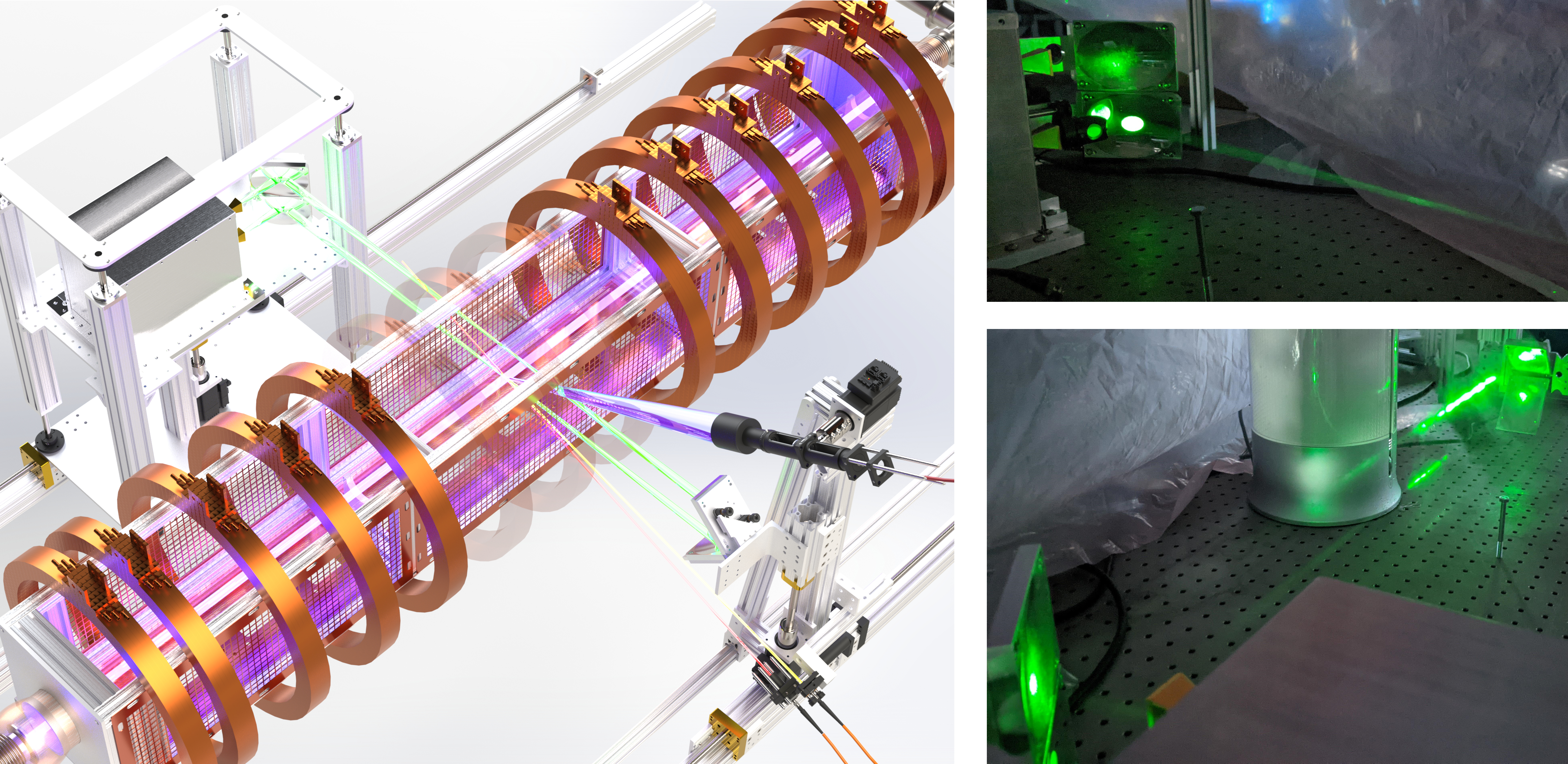}
		\caption{Left: CAD render of the interferometer beam, represented in green, and complete mirror system. The beam passes through MAP's magnet cage, Faraday screen, and plasma before being sent back into the interferometer enclosure 76 mm above the outgoing beam. Both the source side on the right and return side on the left are height adjustable as explained later in \cref{sec:motionSystem}. Both halves of the interferometer are mounted on two-axis computer-controlled motion platforms to allow vertical and horizontal travel along the guidance rails. In addition to the return mirror system, the right-hand platform has mounts for a laser-induced fluorescence system (LIF), (red and blue beams), and passive spectroscopy, (yellow line-of sight). A detailed description of these other diagnostics can be found in \cite{granetzny2025MAP}. Upper right: Beam path visualization and mirror performance check on a test-bench using the alignment laser scattered off water vapor. The lower elliptical mirror focuses the beam to a narrow width to provide high-spatial resolution in the MAP plasma core. Lower right: The beam stays well-focused throughout the entire optical system before returning to the receiving mirror on the left. Without the elliptical mirror system, the beam width at the receiver would be 34 cm, wider than the entire interferometer enclosure in \cref{fig:hetInfOverviewReal}.}
		\label{fig:mirrorSystemCAD}
	\end{center}	
\end{figure*}

\paragraph{Quasi-optics}
Our quasioptical design uses one planar and three elliptical mirrors as shown in \cref{fig:mirrorSystemCAD}. The approximately Gaussian probe beam originates from the lower horn antenna and has its beam waist at the apex of the horn. The first elliptical mirror is designed and positioned such that this apex lies at its first focal point. The second focal point lies on the axis of the vacuum vessel to minimize the probe beam waist in the plasma core. After the beam passes out of the vacuum vessel, it encounters a planar mirror that deflects it upwards into a second elliptical mirror. The latter's first focal point is theoretically below the planar mirror. However, due to right-angle deflection at the planar mirror, its real focal point is inside the plasma. Its second focal point is 7.62 cm above the plasma focal point and sends the probe beam back towards the interferometer enclosure a few centimeters above the vacuum vessel. The beam then reaches a fourth and final elliptical mirror that focuses it into the receiving upper horn antenna. All four mirrors were machined out of 6061 aluminum on a 3-axis CNC mill. The mirror surfaces were polished with increasingly finer grit papers and buffing compound, yielding an almost optical grade finish. Unlike in a waveguide, this optical system does not attenuate the probe beam, which at 105 GHz in a WR-10 waveguide would be 7 dB over the 2.5 m long probe beam path\cite{waveguide}. The optical system allows us to split the setup into a source and return side that can be moved independently as shown in \cref{fig:mirrorSystemCAD} and previously in \cref{fig:MAPTop}. This allows us to move the system to any axial location despite the magnets, power cables and structural supports that would prevent easy repositioning when using a rigid waveguide-based return system.\\

\paragraph{Alignment}
    All four mirrors are set on two-axis kinematic mounts and the system could in principle be aligned by adjusting all four mirrors while only looking at the signal strength of the measurement signal output on the PCB described in \cref{sec:circuit}. However, this procedure would be extremely time-consuming and it would be difficult to establish a rough alignment as a starting point. To simplify alignment we set up an optical laser to mimic the beam shape and source point of the microwave beam. This setup was shown previously in \cref{fig:hetInfOverviewReal}. We start with a well-collimated laser beam and send it through a plano-convex lens mounted inside a lens tube coaxial with the laser. A flip mirror allows us to couple the laser into the microwave beam path. Meanwhile, we set the laser's focal point at the correct distance from the first elliptical mirror by adjusting the position of the lens inside the lens tube. The result is an optical alignment laser beam with the same wide divergence and apparent origin point as the microwave beam. This setup allows for easy alignment of the entire quasi-optical system while also allowing us to confirm that the elliptical mirrors work as designed.\\

\subsection{Motion System and Interfacing}
\label{sec:motionSystem}

A single interferometer measurement can only provide line-integrated density. However, taking measurements at different heights between the plasma's midplane and radial boundary, allows us to calculate the radial profile shape through techniques such as Abel inversion. MAP's interferometer uses motor-driven two-axis motion platforms to position the system at different heights and axial locations. Two stepper motors move the large platform with the interferometer enclosure on the left in \cref{fig:mirrorSystemCAD}, while two others move the compact return mirror system on the right in \cref{fig:mirrorSystemCAD}. These motors have an integrated encoder and are controlled over ethernet using a modular Python interface that can be linked to the RF generator and the interferometer communication interfaces. This setup enables automated measurement campaigns, for example sampling different axial and height positions at different RF power levels. Moreover, this system ensures that the beam path stays precisely aligned when moving the two halves of the interferometer to different measurement locations. 


\subsection{Measurement Performance}

\begin{figure}[htp]
	\begin{center}
		\includegraphics[width=1.0\linewidth]{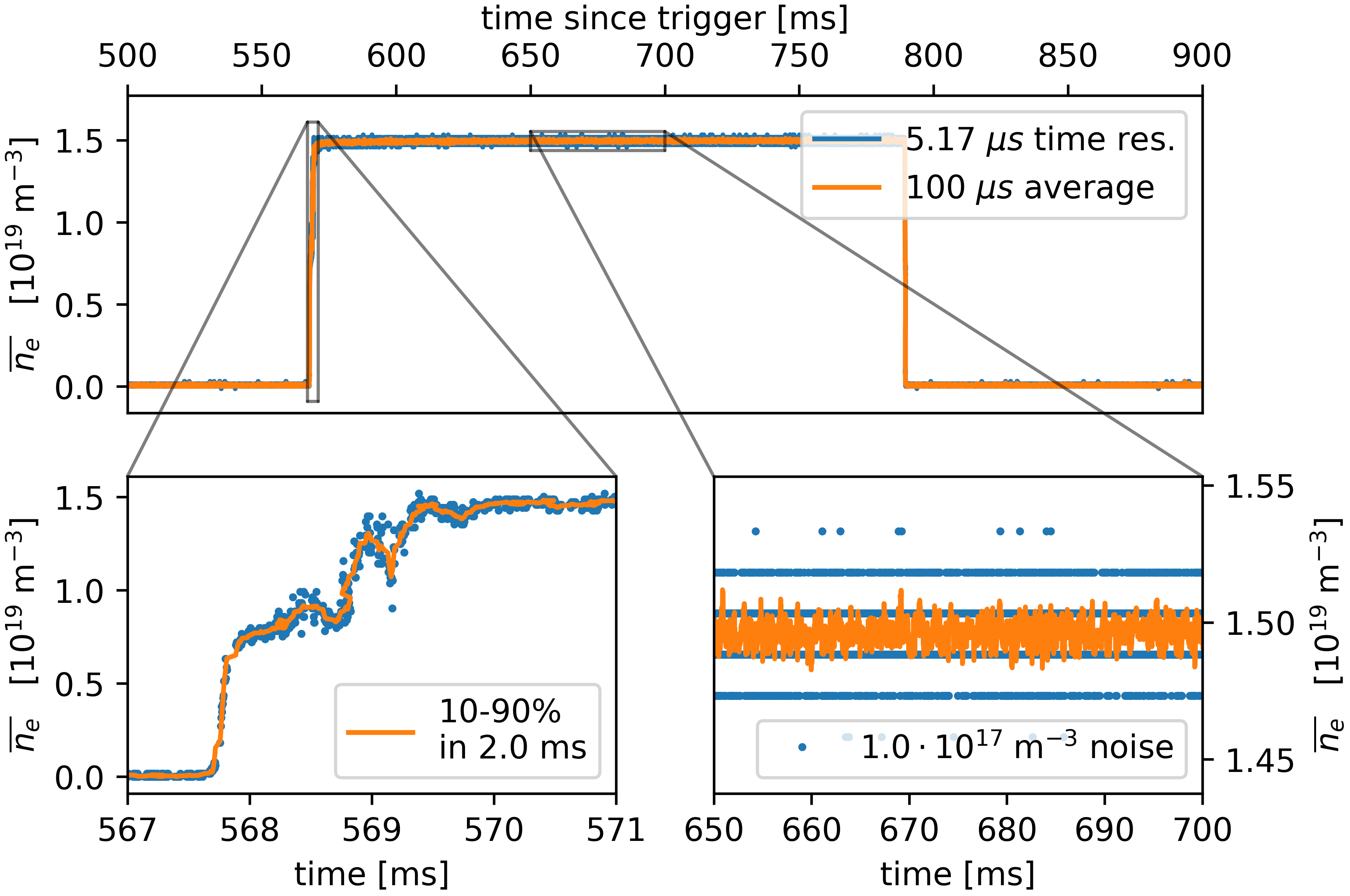}
		\caption{Interferometer density measurement during a short plasma pulse in MAP at 1.3 kW of RF power. The blue data points show the live-streamed data, with a time-resolution of $5\U{\mu s}$. The orange curve shows a $100\U{\mu s}$ moving average. The line averaged-density $\overline{n_e}$ reaches $1.5\ed{19}$ with a 10-90\% rise time of 2 ms. In a shot such as this, the incoming probe beam signal typically has an amplitude of a few hundred $\mu V$, when reaching the mixed signal circuit. The circuit filters out the significant RF noise and auto-adjusts the amplifier gain pre-shot to provide a signal in the ideal voltage needed for clean switching at the high-speed comparator. The result is a noise level of $1.0\ed{17}$, or $0.7\%$, during the steady-state period.}
		\label{fig:hetInfTestPlasma}
	\end{center}	
\end{figure}

Using the modular control interface allows us to trigger the RF plasma discharge and interferometer measurement at the same time. An example of this is shown in \cref{fig:hetInfTestPlasma}, where the discharge occurs 570 ms after the control computer sends the trigger signal to the RF generator and interferometer. After a 10-90\% rise time of 2 ms, the line-averaged density is steady at $1.5\ed{19}$ during the 200 ms long pulse and has a standard deviation of $1.0\ed{17}$. This noise level represents $0.7\%$ of the absolute density value and is below the density resolution of one hundredths of a fringe, equivalent to $3.6\dg$ or a line-averaged density of $1.5\ed{17}$. Measurements such as \cref{fig:hetInfTestPlasma} are the basis for absolute calibrations of MAP's laser-induced fluorescence system \cite{granetzny2025MAP}, that allows us to conduct steady-state density measurements beyond the interferometer cut-off and into the $\ebd{20}$ range. This process is explained in detail in \cite{granetzny2025MAP}.

\section{Summary}

We have developed a new heterodyne microwave interferometer that can perform high-precision, high-speed, real-time plasma density measurements in an environment with significant RF noise. Our design uses a single microwave source together with a power splitter and upconverter to create the probe and reference beams. This setup avoids frequency drift issues and obviates the need for a second high-stability source and master oscillator. The interferometer uses elliptical mirrors to focus the beam into the plasma and guide it back into the receiver. This mirror system is split into a source and return side that can move independently using computer-controlled two-axis motion platforms. This setup enables fast repositioning of the interferometer to measure at any axial location on MAP helicon plasma source despite the magnets, wiring and structural supports that would prevent repositioning of a waveguide-based beam return system. The interferometer is tied into MAP's modular control interface for diagnostic positioning, RF plasma pulsing, interferometer triggering and readout, which allows automation and scripting of measurement campaigns. The measurement signal is amplified, RF noise-filtered and analyzed directly in the interferometer enclosure using a high-speed, high-precision mixed signal circuit and an off-the-shelf FPGA. Our design can measure a one hundredth of a fringe and provides line-integrated density measurements from $1.5\ed{17}$ into the $\ebd{19}$ range. Measurements have an uncertainty of $1.0\ed{17}$ and are received in real time every $5\U{\mu s}$ in a compact and flexible setup that does not require a high-speed digitizer or oscilloscope. 

\section*{Acknowledgements}
The research presented here was funded by the National Science Foundation under grants PHY-1903316 and PHY-2308846 as well as the College of Engineering at UW-Madison.



\appendix
\setcounter{equation}{0}
\renewcommand{\theequation}{A\arabic{equation}}

\section{Designing an Elliptical Mirror}
\label{app:MirrorDesign}

An elliptical mirror exploits the fact that in an ellipse any ray originating from one focal point will be reflected into the other. This is demonstrated in \cref{fig:mirrorDesign}, for an ellipse with major radius $a$, minor radius $b$, and focal points at $x_{focus} = \pm c$. A beam launched from the left focal point moves a distance $d_1$, is reflected an angle $\delta$, and reaches the second focal point after a distance $d_2$. For example, in case of the first elliptical mirror in our interferometer, $d_1$ is the distance from the probe beam source to the mirror surface, $d_2$ is the distance from the mirror to the plasma core and $\delta$ is $90\dg$. If we rotate the ellipse in \cref{fig:mirrorDesign} around the x- or y-axis, the resulting body is an ellipsoid that has the same focusing characteristics as the original ellipse but can be used with a real three-dimensional beam.\\

\pic[1.0]{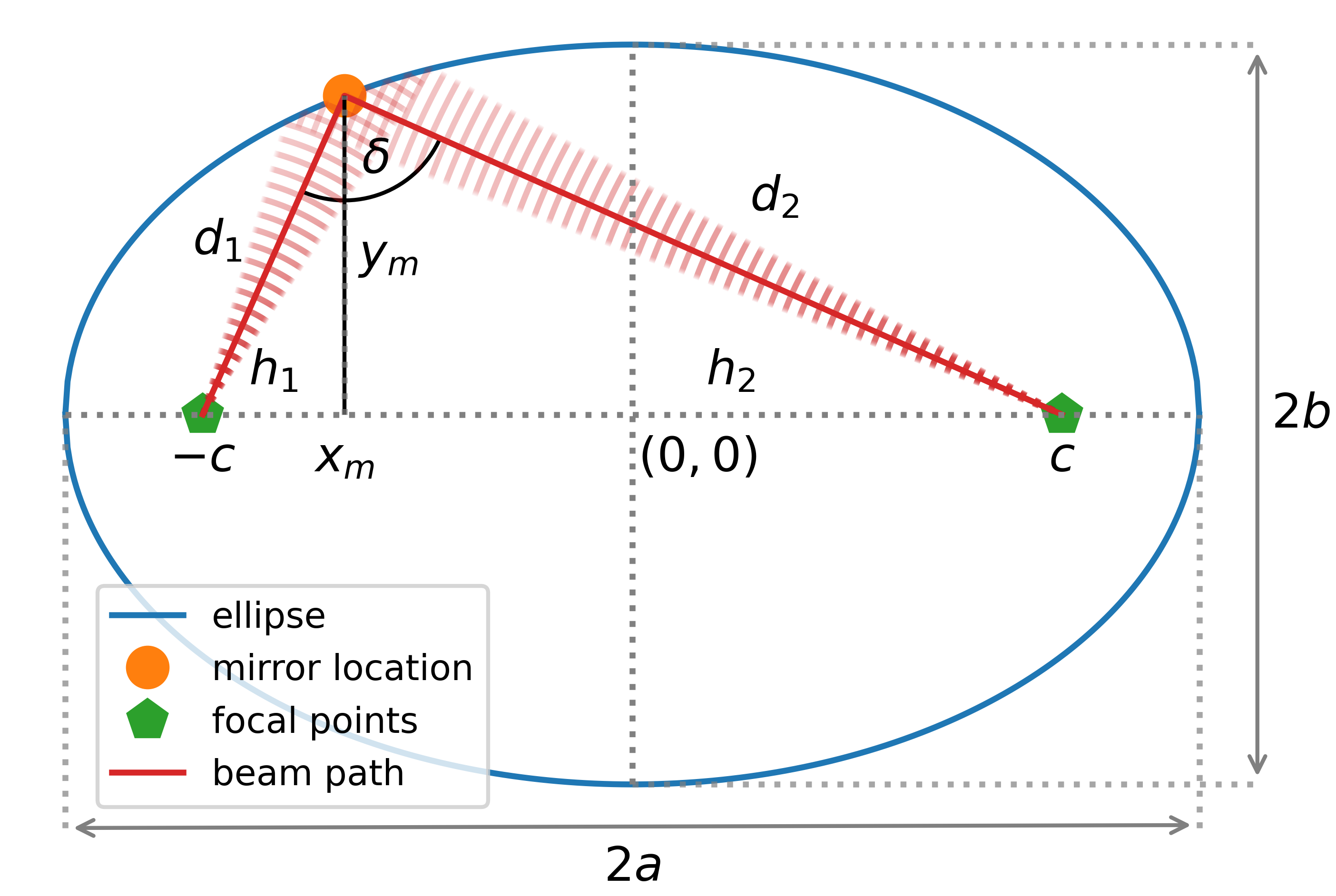}{Beam reflection in an ellipse and needed measurements of major and minor axis, $a$ and $b$, to calculate the parameters defining an elliptical mirror with for a given pair of focal lengths $d_1$ and $d_2$.}{fig:mirrorDesign}

\noindent
The equations defining an ellipse can be expressed as

\begin{align}
    y^2 &= b^2 \brk{1 - \frac{x^2}{a^2}}\label{eq:ellipseY}\\
    b^2 &= a^2 - c^2\label{eq:mirrorBSol}.
\end{align}

Our task is to find $a$ and $b$ such that they define an ellipse with the needed reflection characteristics. In practice, we do not need the full ellipse but only the section that intersects our beam. We therefore need to find the intercept coordinates $x_m$ and $y_m$. By examining \cref{fig:mirrorDesign}, we find the following useful relations:

\begin{align}    
    d_1^2 &= h_1^2 + y_m^2\label{eq:pyth1}\\
    d_2^2 &= h_2^2 + y_m^2    \label{eq:pyth2}\\    
    h_1 &= c + x_m\label{eq:lin1}\\    
    h_2 &= c - x_m\label{eq:lin2}\\
    \brk{2c}^2 &= d_1^2 + d_2^2 - 2 d_1 d_2 \cos\delta\label{eq:pyth3}.
\end{align}

\noindent
\Cref{eq:pyth3} immediately yields
\begin{align}
 c &= \frac{\sqrt{d_1^2 + d_2^2 - 2 d_1 d_2 \cos\delta }}{2}\label{eq:mirrorCSol}.
\end{align}

\noindent
By adding and subtracting \cref{eq:pyth1,eq:pyth2} and substituting in \cref{eq:lin1,eq:lin2} we get
\begin{align}
d_1^2 + d_2^2 &= \brk{c+x_m}^2 + \brk{c-x_m}^2 + 2y_m^2\label{eq:mirror1},\\
d_1^2 - d_2^2 &= \brk{c+x_m}^2 - \brk{c-x_m}^2\label{eq:mirror2}.
\end{align}

\noindent
\Cref{eq:mirror2} can directly be solved for $x_m$, yielding:
\begin{align}
    x_m = \frac{d_1^2 - d_2^2}{4c}\label{eq:xmSolve}
\end{align}

\noindent
We can solve for $a$ by substituting \cref{eq:mirrorBSol,eq:ellipseY,eq:xmSolve} into \cref{eq:mirror1} and find after a small calculation

\begin{align}
   0 &= a^4 - a^2\frac{d_1^2 + d_2^2}{2} + \brk{\frac{d_1^2 - d_2^2}{4}}^2\label{eq:five}
\end{align}

\noindent
\Cref{eq:five} has only one positive real solution for $a$ under the condition $a > c$, namely

\begin{align}
    a &= \frac{d_1 + d_2}{2}.\label{eq:mirrorASolve}
\end{align}

\Cref{eq:mirrorASolve,eq:mirrorBSol,eq:mirrorCSol} fully define the ellipse. We can use this solution to create a CAD model for the elliptical mirror by defining the ellipse's curve analytically and rotating it to build the corresponding ellipsoid. After cutting out the needed section using the intercept coordinates in \cref{eq:ellipseY,eq:xmSolve}, we are left with a CAD model of the elliptical mirror suitable for manufacturing using three-axis machining, 3D-printing or other techniques.\\

\section*{References}
\bibliography{references}

\end{document}